\documentclass[twocolumn,secnumarabic,amssymb, floatfix, nobibnotes, aps, prl, superscriptaddress]{revtex4-1}
\pdfoutput=1

\usepackage{float}
\usepackage{graphicx}
\usepackage{amsmath} 
\usepackage{epstopdf}
\usepackage{setspace}
\usepackage[shortlabels]{enumitem}
\usepackage{mathrsfs}
\usepackage{color,soul}
\usepackage[dvipsnames]{xcolor}
\usepackage{dcolumn,amssymb}
\usepackage{comment}
\usepackage[normalem]{ulem}

\usepackage{dcolumn} 
\usepackage{natbib}
\usepackage{bm} 
\usepackage{float} 
\usepackage{bbm} 
\usepackage{color}
\usepackage{amsfonts}
\usepackage{lmodern} 
\usepackage{amsmath,amssymb}

\usepackage[colorlinks=true,citecolor=blue]{hyperref}
\hypersetup{colorlinks=true,citecolor=blue,linkcolor=red,urlcolor=blue}

\begin{document}
\title{Active extensile stress promotes 3D director orientations and flows}
\author{Mehrana R. Nejad}
\author{Julia M. Yeomans}
\affiliation{The Rudolf Peierls Centre for Theoretical Physics, Department of Physics, University of Oxford, Parks Road, Oxford OX1 3PU, United Kingdom}
\begin{abstract}
We use numerical simulations and linear stability analysis to study an active nematic layer where the director is allowed to point out of the plane. Our results highlight the difference between extensile and contractile systems.
Contractile stress suppresses the flows perpendicular to the layer and favours in-plane orientations of the director. By contrast extensile stress promotes instabilities that can turn the director out of the plane, leaving behind a population of distinct, in-plane regions that continually elongate and divide. This supports extensile forces as a mechanism for the initial stages of layer formation in living systems, and we show that a planar drop with extensile (contractile) activity grows into three dimensions (remains in two dimensions). The results also
explain the propensity of disclination lines in three dimensional active nematics to be of twist-type in extensile or wedge-type in contractile materials. 
\end{abstract}
\maketitle


\begin{figure*}[t] 
    \centering
    \includegraphics[width=0.99\textwidth]{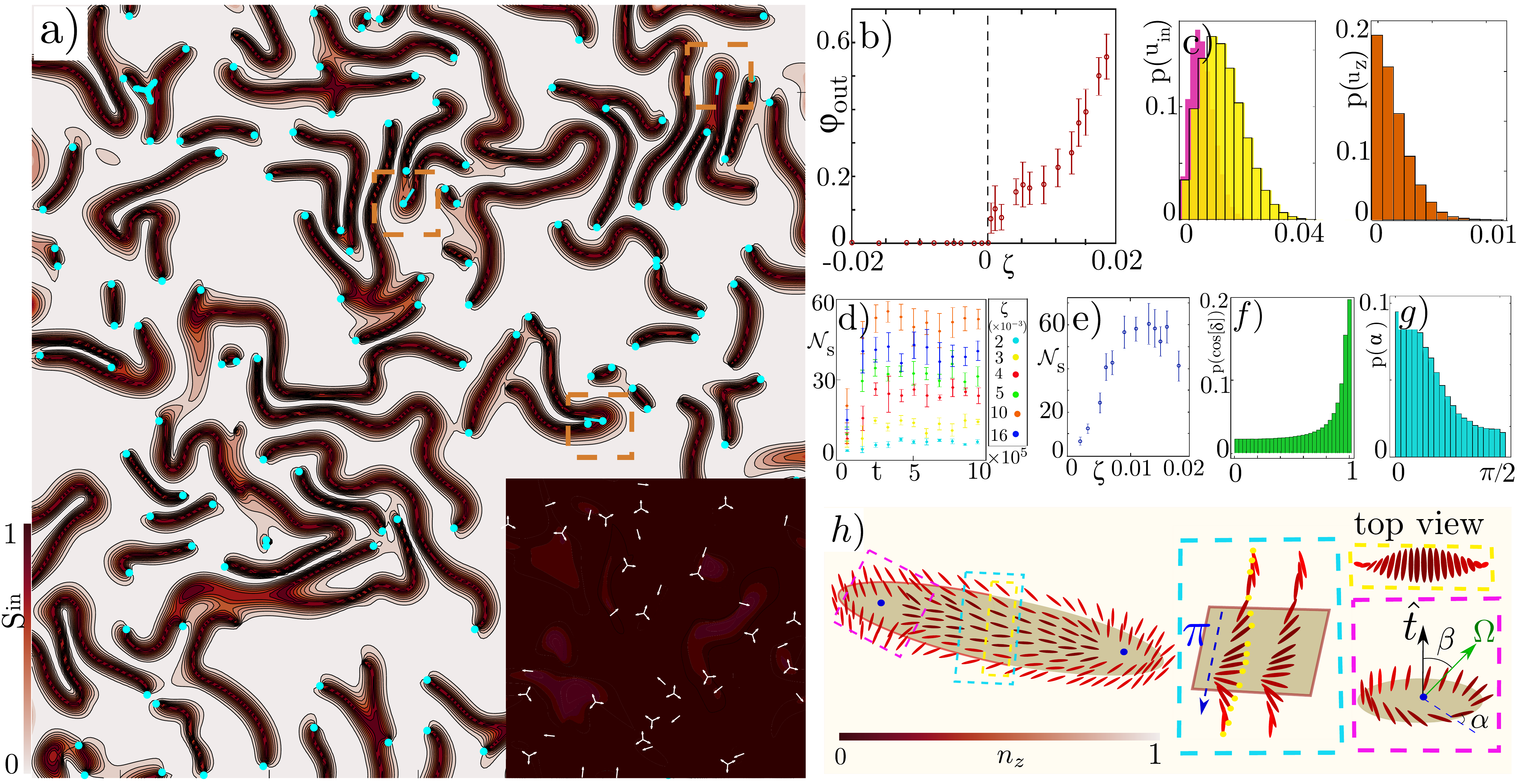}
    \caption{Snapshots from simulations of 2D active nematics layers. Color denotes the magnitude of the in-plane order ($S_{in}=[1-n_z^2]^{1/2}$) from in-plane (black) to out-of-plane (white): a) extensile stress, defects in cyan. 3D twist-type defects are represented by circles. Inset: contractile stress, 2D $\pm 1/2$ topological defects are shown in white. b) Area fraction of regions with out of plane director as a function of activity.
    c) In-plane, $u_{\text{in}}$, and out-of-plane, $u_{\text{z}}$, speed for contractile (yellow, $u_{\text{z}}=0$) and extensile (pink, orange) driving. d) Variation of the number of snakes $\mathcal{N}_s$ with time. e) Number of snakes as a function of activity at steady state.
     f) Distribution of the angle between the director and the tangent to the boundary of snakes. The histogram peaks at $\cos(\delta)=1$, indicating parallel alignment of the director with the boundary.  
     g) Distribution of the angle $\alpha$ in extensile systems.
      h) Director in a snake. Cyan outline: 
    moving along the blue dashed arrow, crossing the width of a snake, the director twists by $\pi$. Yellow outline:  top view of the director across the width of a snake. Magenta outline: 3D twist type defects are commonly found at the ends of the snakes. The blue circle shows the core of the defect, $\hat{t}$ represents the normal to the layer, and the blue dashed line connects the center of the twist type defect to the position with an in-plane director. $\alpha$ is the angle between this line and the local director. The director rotates out of the plane around the rotation vector $\Omega$ to form the twist defect.\label{fig:wet_system}}
\end{figure*}

Because living systems exist out of equilibrium~\cite{Ramaswamy_2017} they can exhibit novel behaviors that cannot be captured by
conventional equilibrium statistical mechanics ~\cite{RevModPhys.85.1143} such as coherent animal flocks \cite{flocking} and cell crawling and division \cite{PhysRevLett.123.118101,doi:10.1146/annurev.biochem.73.011303.073844}.   

 \begin{figure}[t] 
    \centering
    \includegraphics[width=0.45\textwidth]{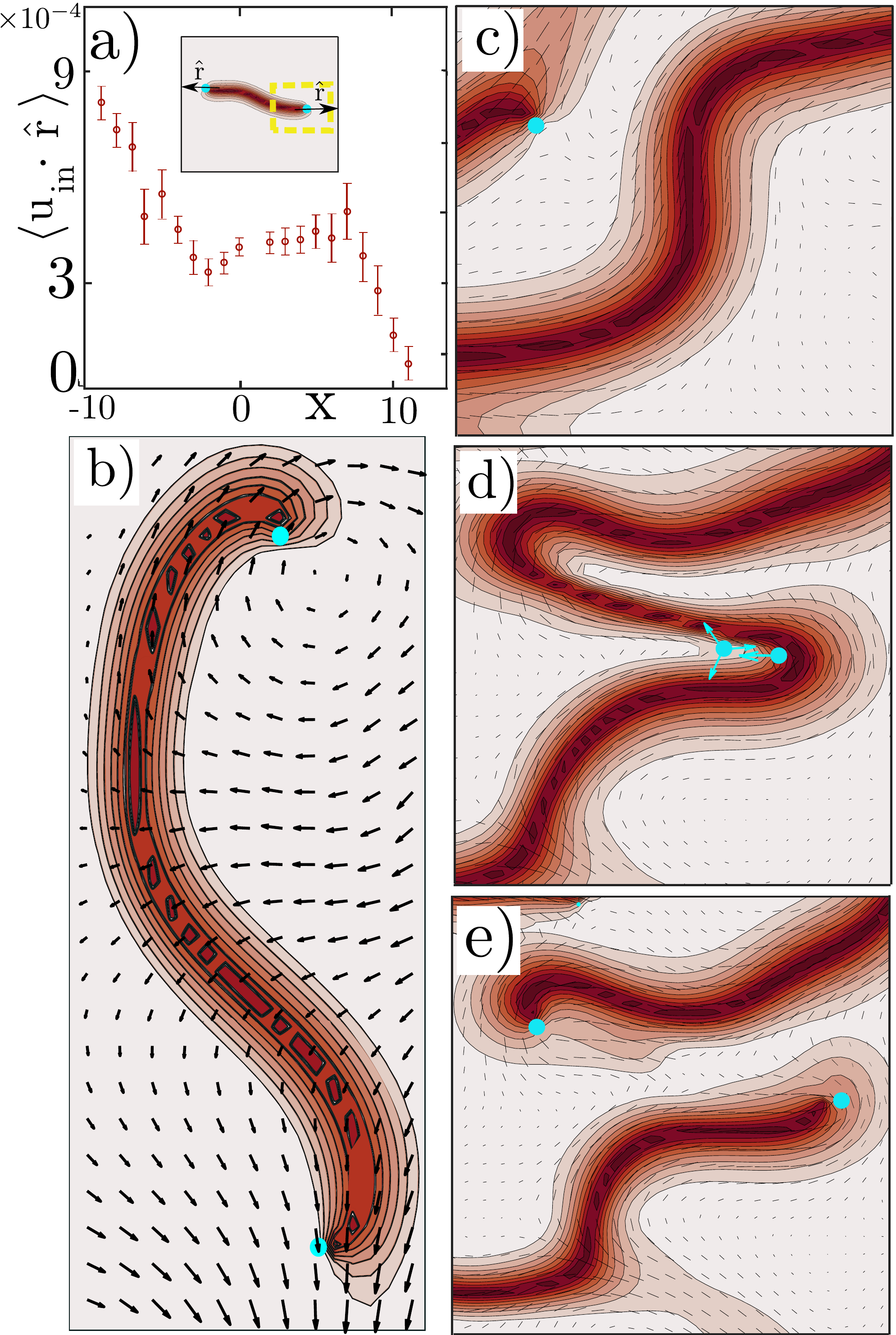}
    \caption{a) Average flow along the direction of the elongation of snakes (measured at the ends, i.e., within the yellow box). The origin corresponds to the blue dot and $x>0$ always corresponds to outside the snake. $\langle \textbf{u}_{in} \cdot \hat{r} \rangle >0$, indicating that active flows elongate the snakes. b) Flows around a snake. c)-e) Evolution of a snake. Elongated snakes undergo a bend instability and form defects. The director then moves to the third direction dividing the snake into two with twist-type defects at their ends.}
\label{fig:3}
\end{figure}
 The hydrodynamics of active particles, such as bacteria, cells and microtubules driven by kinesin motors lies in the low Reynolds number regime and they can be modeled as force dipoles with nematic symmetry. The active stress is then proportional to the nematic tensor $\textbf{Q}$, and can be written as $-\zeta \textbf{Q}$, where $\zeta$ is the activity. The direction of the force dipoles can be inward along the dipolar axis (contractile systems, $\zeta<0$) or outward (extensile systems, $\zeta>0$). Both extensile and contractile systems are found in nature: actomyosin suspensions are contractile, microtubule-kinesin motor suspensions and bacteria are extensile, while confluent cell layers can be either \cite{Balasubramaniam2021,duclos2018spontaneous}.
 
Active stresses destabilise the nematic phase resulting in a state of chaotic flows, with prominent vorticity and fluid jets, known as active turbulence \cite{fraden2019turbulent,PhysRevLett.110.228102,Wensink14308,slomka2017spontaneous,martinez2019selection,PhysRevLett.120.208101,lemma2019statistical,thampi2016active}. In two dimensions active turbulence is characterized by the continuous creation and annihilation of $+1/2$ and $-1/2$ topological defect pairs in the nematic director field \cite{cortese2018pair,schaller2013topological,PhysRevLett.125.218004}. The $+1/2$ defects have polar symmetry and hence are self-propelled.
Recently, three-dimensional active suspensions have also attracted a considerable amount of interest. In contrast to 2D, 3D active flows are governed by the creation and annihilation of disclination lines and loops ~\cite{duclos2020topological,PhysRevLett.125.047801,PhysRevLett.124.088001}.

There is significant understanding of completely 2D or 3D active materials. However many biological systems evolve from 2D to 3D structures during their life cycle. The growth of a 2D layer into the third dimension leads to the formation of biofilms (due to pushing forces that the growing cells exert on their surroundings) \cite{hartmann2019emergence,you2019,chu2018self,duvernoy2018asymmetric,grobas2021swarming,doi:10.1080/1539445X.2021.1887220}, 
where the transition can be initiated by the formation of a vertically aligned core of bacteria 
 \cite{warren2019spatiotemporal,beroz2018verticalization}. Gastrulation is a vital step in the early development of most animals when a single layer epithelium is reorganised into a multilayer structure of differentiated cells that will form specific tissues and organs \cite{Krezel2020}.
To understand these processes further, in this Letter we study the transition from 2D to 3D in active nematics showing that extensile and contractile materials demonstrate remarkably different behaviours. 

We first consider a 2D active nematic system in the $x$-$y$ plane. Allowing the director and velocity fields to have components along $x$, $y$ and $z$ directions (and starting the simulations with small 3D noise), we numerically solve the active nematohydrodynamic equations of motion. This is now well-documented ~\cite{PhysRevE.76.031921,PhysRevE.76.031921,B908659P,PhysRevLett.126.227801,Carenza2019,C6SM00812G}, and we present the equations and simulation details in Supplemental Material (SM).

\noindent
{\bf {Results:}}
\label{pair-wise}
In Fig.~\ref{fig:wet_system}  we compare the behaviour of an extensile system with activity $\zeta=0.008$ with that of a contractile system with $\zeta=-0.008$. The figure shows that contractile stresses suppress perturbations of the director field in the direction perpendicular to the layer, leading to the usual 2D active dynamics (Fig.~\ref{fig:wet_system}(a), inset). In extensile systems, however, the director has non-zero out-of-plane components except within dynamic, elongated domains (Fig.~\ref{fig:wet_system}(a) and Movie 1). 
This behaviour is quantified in Fig.~\ref{fig:wet_system}(b) where we plot the area fraction of the out-of-plane regions as a function of activity, showing that this quantity remains zero for the contractile case, but increases with activity in extensile systems.  Histograms of the corresponding in-plane and out-of-plane flow fields are shown in Fig.~\ref{fig:wet_system}(c).
In contractile systems flows remain in the $x$-$y$ plane whereas in the extensile case the flow develops substantial components along $z$ which act to drive the director into the third dimension.

We will discuss the dynamics of the in-plane domains in the extensile case below, but first we 
perform a linear stability analysis of the nematohydrodynamic equations around the fully-aligned in-plane nematic phase to further understand the different behaviour of extensile and contractile systems. 
Decomposing any fluctuating field as $\delta f(\textbf{r},t)=\int d \textbf{q}\: d \omega \tilde{f}(\textbf{q},\omega)\: e^{i \textbf{q} \cdot \textbf{r} +\omega t}$, in the long-wave-length and zero Reynolds number limit the growth rate of a perturbation reads
\begin{equation}\label{lsa}
     \omega_{out}= \frac{ 3\zeta }{4\eta}\cos^2 \theta, \:\: \omega_{in}= \frac{3 \zeta}{4\eta} \cos 2 \theta,
\end{equation}
where $\omega_{out}$ and $\omega_{in}$ are the growth rates of the out-of-plane ($\delta Q_{xz}$) and in-plane ($\delta Q_{xy}$) components of the director (which are decoupled eigenmodes), $\theta$ is the angle between the wavevector of the perturbation and the direction of the order and $\eta$ is the viscosity. 
The full form of the growth rates (including the flow-aligning dependence) can be found in SM. Here we have set the flow-aligning parameter to zero.
\begin{figure}[b]
    \centering
    \includegraphics[width=\linewidth]{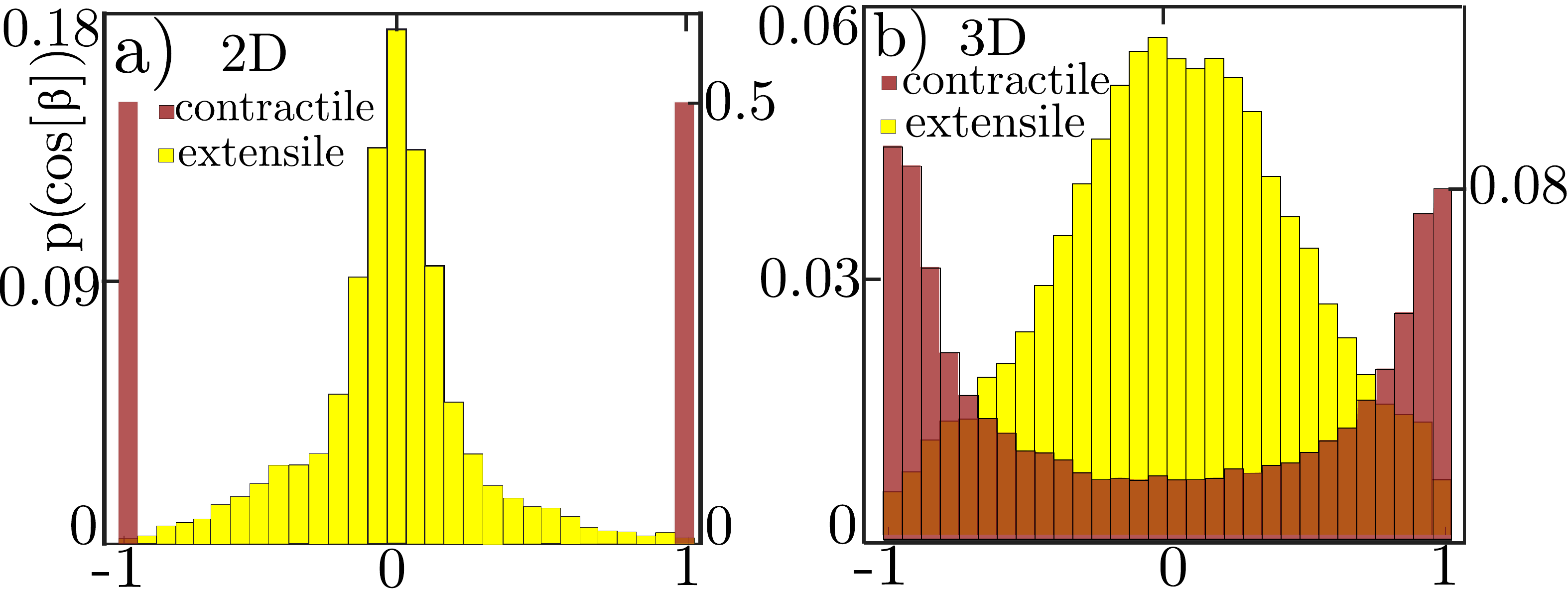}
    \caption{Twist angle $\beta$ in (a) 2D, (b) 3D. In extensile system most defects are twist-type ($\beta=\pi/2$) while in contractile systems most defects are wedge-type ($\beta=0,\pi$). Right/left axis shows $p(\cos[\beta])$ in the contractile/extensile system.}
    \label{fig:fig3a}
\end{figure}

\begin{figure*}[t]
    \centering
    \includegraphics[width=\linewidth]{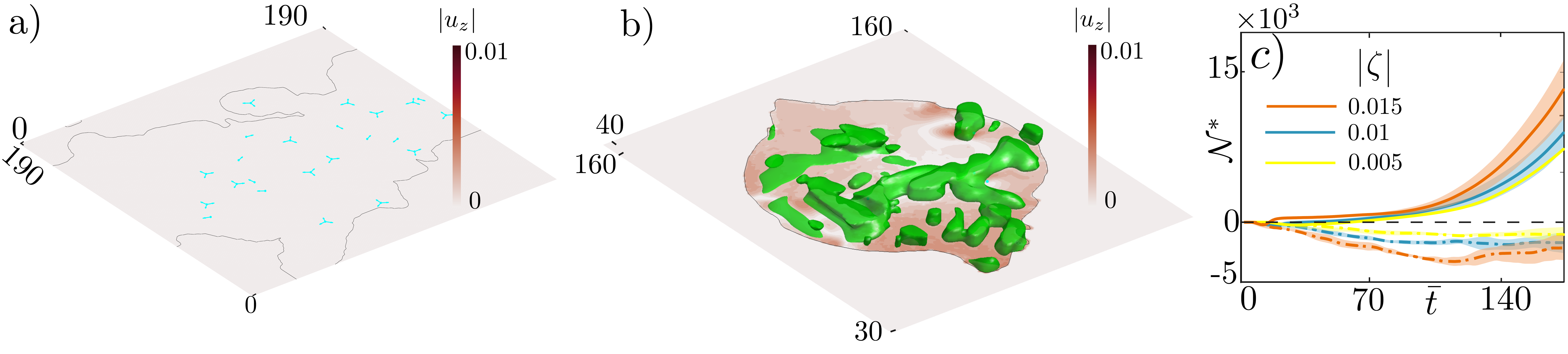}
    \caption{3D views of a growing drop with (a) contractile, (b) extensile activities. Defects are represented in cyan, green shows the out-of-plane concentration and the red shading indicates out-of-plane speed. (Note (b) is a close-up). (c) The difference in the number of out-of-plane particles in an active growing system ($\zeta\neq 0$) and in a passive growing system ($\zeta = 0$) for extensile (solid lines) and contractile (dashed-dotted lines) activity. The semi-transparent colors show error bars calculated using the standard deviation from five simulations.}
    \label{fig:fig4}
\end{figure*} 

The onset of the instability occurs when $\omega_in$ or $\omega_out$ become positive. The in-plane component shows the well-known instability of 2D active nematics to bend (splay) perturbations in extensile (contractile) systems \cite{Sriram2002}. By contrast the growth rate of the out-of-plane component does not change sign for different values of $\theta$ and is positive (negative) for extensile (contractile) systems. Thus the out-of-plane component only grows in extensile systems. 
 \noindent
 
 {\bf In-plane domains in extensile active nematics:} We now return to consider the elongated domains where the director field remains in the $x$-$y$ plane. These initially form in regions of splay distortion where the flow fields along the $z$-axis are insufficiently strong to push the director field out of the plane. For convenience we will refer to the in-plane domains as {\em snakes}.
  
After a transient phase the average number of snakes fluctuates around a constant value (Fig.~\ref{fig:wet_system}(d)) 
 which first increases and then decreases with increasing activity (Fig.~\ref{fig:wet_system}(e)).
This reflects a balance between snakes elongating and then breaking up, 
and the possibility that local flows will be sufficiently strong to destroy a snake by pushing the director out of the $x$-$y$ plane.

To understand how the snakes lengthen and divide we study their director and flow fields. 
We first measure $\cos \delta$, the angle between
 the orientation of the director at the boundary of a snake and the local tangent to the boundary.
The histogram of $\cos \delta$, shown in Fig.~\ref{fig:wet_system}(f)
peaks at $\cos \delta=1$, showing that the director inside a snake tends to align parallel to its elongation direction.

Therefore, along the boundaries of a snake the director rotates through $\pi/2$ to match the surrounding vertical director configuration (Fig.~\ref{fig:wet_system}(h)) and as a result, looking at a cross-section across the width of a snake, the director twists by $\pi$ (cyan and yellow outlines in Fig.~\ref{fig:wet_system}(h)). At the ends of the snake this results in twist defects with the director configuration shown in the magenta outline in Fig.~\ref{fig:wet_system}(h).

An angle which characterises twist defects is $\alpha$, the angle between the line  which connects the center of the defect to the position of the in-plane director and the  direction of the in-plane director itself (magenta outline in Fig.~\ref{fig:wet_system}(h)).  
It has been shown that { tangential} twist is more predominant in extensile defect { loops} \cite{PhysRevLett.124.088001}. However, for the point-like defects in our quasi-2d system,
Fig.~\ref{fig:wet_system}(g) shows that $\alpha$ has a peak at $\alpha=0$, indicating that the twist-type defects are predominantly radial.

We are now in a position to understand the dynamics of a snake.
The stresses which result from the twist defects set up flows which act to elongate the snakes and to align the director field inside them (Figs.~\ref{fig:3}(a) and ~\ref{fig:3}(b)). 
The ensuing evolution is illustrated in Figs.~\ref{fig:3}(c)-(e).
Since the system is extensile and the director is parallel to their length, the elongated snakes undergo a bend instability (see Movie 1). The growth of a bend deformation is equivalent to the formation of a pair of two-dimensional, $\pm 1/2$ defects (orange outlines in Fig.~\ref{fig:wet_system}(a) and Fig.~\ref{fig:3}(c)-(e)).  
We have seen that bend deformations are unstable to  director perturbations perpendicular to the layer and, due to the large bend deformations at the position of the defects, the director rotates out of the plane and the snake splits into two smaller snakes terminated by twist director configurations (Fig.~\ref{fig:wet_system}(h)).

\noindent
{\bf Relation to 3D active turbulence:} 
The behavior that we have identified in two dimensions persists into three dimensions. Fully-developed 3D active turbulence is characterised by motile disclination lines that form closed loops that can appear, grow, shrink and disappear~\cite{duclos2020topological,PhysRevLett.125.047801,PhysRevLett.124.088001}. Experiments and simulations have shown that twist (wedge) type defects are observed in extensile (contractile) systems ~\cite{duclos2020topological,PhysRevLett.125.047801} which can be explained by our model. The director configuration on a plane locally perpendicular to a disclination line can be characterised by the
twist angle $\beta$ between 
the tangent to the disclination line and the axis around which the director rotates out of the plane (Fig.~\ref{fig:wet_system}(h)).  $\beta = 0, \pi$ 
correspond to cross sections of the disclination line with the configurations of $+1/2$ and $-1/2$ defects, respectively, and other values indicate degrees of twist with $\beta = \pi/2$
corresponding to a pure twist configuration.
Figures ~\ref{fig:fig3a}(a) and ~\ref{fig:fig3a}(b) compare the distribution of $\beta$ for the single layer considered here to simulations of full 3D active turbulence keeping the activity the same, and considering both extensile and contractile systems. 
The figure shows that similar behaviour is observed in both 2D and 3D: in contractile systems defects are predominantly two dimensional, whereas in the extensile system there is a clear preference for introducing twist defects. We show in SM that the growth rate of twist perturbations in a fully 3D active nematic is also given by $\omega_{out}$, Eq.~(\ref{lsa}), confirming that twist perturbations grow (get suppressed) in extensile (contractile) systems. As a result, at linear order, twist-type disclinations are only allowed to form in extensile systems.

We have also run simulations on 3D systems confined in one of the dimensions by free-slip boundaries with parallel anchoring (the choices are motivated by experiments on bacteria and eukaryotic cells where cells can move along the substrate) to check that the difference between extensile and contractile systems leads to 
promotion of flows perpendicular (parallel) to the boundaries, due to the formation of twist- (wedge-) type defects in extensile (contractile) systems. This is in agreement with the growth of twist observed in extensile systems in channels \cite{PhysRevLett.125.257801}.

{\bf Active growing droplets:} To more explicitly investigate the role of active extensile and contractile flows on 3D growth, we performed 3D simulations on an initially planar, $2D$ growing drop embedded in a 3D passive fluid. The growth is added by increasing the concentration of active material by a rate $k_d$ (see Supplemental Material \cite{sm}). 
3D views of the drop at $\bar{t} = t \zeta/\eta=210$ (where $\eta$ is viscosity) are shown in Figs.~\ref{fig:fig4}(a) and (b) for contractile and extensile systems, respectively. The black outline shows the boundary of the drop in the initial 2D plane and the out-of-plane concentration is represented in green \cite{sm}. The red background color shows the speed perpendicular to the layer, $|u_z|$. In the contractile system the out-of-layer velocity is negligible and the growth is {mainly} in the plane.  By contrast, the extensile drop has non-zero velocity perpendicular to the layer and grows into the third dimension (see Movies 2 and 3 for the dynamics). This behaviour is confirmed more quantitatively in Fig.~\ref{fig:fig4}(c) which shows the evolution of $\cal{N}^{*}$,
 the difference between the number of out-of-plane particles in an active growing system and in a passive growing system \cite{sm}. $\cal{N}^{*}$ is positive (negative) in extensile (contractile) systems showing that extensile (contractile) flows promote (suppress) growth to the third direction.\\

To summarize, in an extensile (contractile) system there is a positive (negative) feedback between out-of-plane flows and out-of-plane director orientation and as a result  out-of-plane perturbations grow (are suppressed) in extensile (contractile) systems.
The transition from 2D to 3D is often a vital step in biological processes. Biofilm formation can be initiated by cells turning to point perpendicular to a substrate \cite{warren2019spatiotemporal,beroz2018verticalization,you2019,hartmann2019emergence} which leads to a planar-to-bulk transition \cite{grant2014role}. This has been described in terms of mechanical instabilities due cell divisions which cause extensile stresses \cite{grant2014role,boyer2011buckling,10.1371/journal.pone.0048098} and the mechanism we describe here supports this interpretation.
Moreover epithelial layers fold into 3D configurations during gastrulation, an early stage of embryogenesis; an example is the ingression of cells at the primitive streak in the chick embryo \cite{Nagera2020}. 

The collective motion and stress in epithelial systems can be extensile \cite{saw2017topological,Balasubramaniam2021}, as a result of cell-cell interactions or cell divisions, suggesting that flows out of the plane, known to be important in driving gastrulation, may be related to local extensile activity.%

\section*{Acknowledgements}
M.R.N. acknowledges the support of the Clarendon Fund Scholarships. We thank Amin Doostmohammadi, Liam Ruske, Kristian Thijssen, Rastko Sknepnek and Kees Weijer for helpful conversations.
\bibliographystyle{apsrev4-1}
\bibliography{references}
\end{document}